\documentclass[12pt]{article}

\usepackage[round]{natbib} 
\usepackage{rotating, rotfloat}
\usepackage{graphics, graphicx, hyperref, url, amsmath, amsthm, amsfonts, enumerate, epstopdf, subfigure, booktabs, float, color, caption, txfonts}

\usepackage{listings}
\usepackage{color}

\definecolor{dkgreen}{rgb}{0,0.6,0}
\definecolor{gray}{rgb}{0.5,0.5,0.5}
\definecolor{mauve}{rgb}{0.58,0,0.82}

\hypersetup{
    pdfstartview={FitH},    
    colorlinks=true,        
    linkcolor=blue,         
    citecolor=blue,         
    filecolor=blue,         
    urlcolor=blue           
}

     \def\etal{{\itshape et al.}}
\def\beqn{\begin{eqnarray}} \def\eeqn{\end{eqnarray}} 
\def\endp{\hfill $\square$}

\newcommand{\bec}{\begin{center}}
\newcommand{\enc}{\end{center}}
\newcommand{\bee}{\begin{eqnarray*}}
\newcommand{\ene}{\end{eqnarray*}}
\newcommand{\beq}{\begin{equation}}
\newcommand{\eeq}{\end{equation}}

\begin{document}

\title{\textbf{Fast and exact implementation of 3-dimensional Tukey depth regions}
\thanks{Corresponding author's email: csuliuxh912@gmail.com}}

\author {{Xiaohui Liu$^{a}$}\\
         {\em\footnotesize $^a$ School of Statistics, Jiangxi University of Finance and Economics, Nanchang, Jiangxi 330013, China}\\
}
\date{}
\maketitle

\begin{center}
\begin{minipage} {12cm}{ 
\vspace*{-7mm} \small \textbf{Abstract}.
\small
Tukey depth regions are important notions in nonparametric multivariate data analysis. A $\tau$-th Tukey depth region $\mathcal{D}_{\tau}$ is the set of all points that have at least depth $\tau$. While the Tukey depth regions are easily defined and interpreted as $p$-variate quantiles, their practical applications is impeded by the lack of efficient computational procedures in dimensions with $p > 2$. Feasible algorithms are available, but practically very slow. In this paper we present a new exact algorithm for 3-dimensional data. An efficient implementation is also provided. Data examples indicate that the proposed algorithm runs much faster than the existing ones. 

\vspace{1mm}

{\small {\bf\itshape Key words:} Tukey depth; 3-dimensional Tukey depth regions; Exact algorithm; Fast implementation} \vspace{1mm}

{\small {\bf2000 Mathematics Subject Classification Codes:} 62F10; 62F40; 62F35}}
\end{minipage}
\end{center}

\setlength{\baselineskip}{1.25\baselineskip}

\vskip 0.1 in
\section{Introduction}
\paragraph{}
\vskip 0.1 in \label{Introduction}

Given a data set $\mathcal{X}^{n} = \{X_{1},\, X_{2},\, \cdots,\, X_{n}\}$ in $\mathcal{R}^{p}$, \cite{Tuk1975} proposed to consider the following function
\begin{eqnarray}
\label{HDSV}
d(x,\, F_{n}) = \inf_{u \in \mathcal{S}^{p - 1}} \frac{1}{n} \# \left\{ i: u^{T}x \ge u^{T}X_{i}
,\, \ i \in \mathcal{N} \right\},
\end{eqnarray}
as a tool to measure how central a point $x$ lies in $\mathcal{X}^{n}$, where $F_{n}$ denotes the empirical distribution corresponding to $\mathcal{X}^{n}$,  $\mathcal{S}^{p - 1} = \{v\in R^{p}: \|v\| = 1\}$, $\mathcal{N} = \{1,\, 2,\, \cdots,\, n\}$, and $\# \{\cdot\}$ denotes the number of data points in set $\{\cdot\}$. $d(x,\, F_{n})$ decreases when $x$ moves outwards from the interior of $\mathcal{X}^{n}$, and vanish at $x$ being outside of the convex hull of all observations. Using this, a center-outward ordering can be developed for multivariate observations. Similar to the setting of univariate order statistics, this ordering is affine equivariant, and so are the multivariate estimators constructed on \eqref{HDSV}. To reflect this seminal work of Tukey, \eqref{HDSV} is commonly referred to as Tukey depth (or halfspace depth) in the literature.
\medskip

Being capable to order multivariate observations, Tukey depth usually serves as a convenient way to extend the methods of signs and ranks, order statistics, quantiles, and outlyingness measures to high spaces from their univariate counterparts. Various desirable applications of Tukey depth can be found in the literature; see for example \cite{YS1997}, \cite{LCL2012} and references therein for details. Along the line of \cite{Tuk1975},  many other depth notions have also been proposed in the past decades. Among others, primary are the simplicial depth \citep{Liu1990}, zonoid depth \citep{KM1997}, and projection depth \citep{Liu1992, Zuo2003}. The axiomatic definition of depth functions can be found in \cite{ZS2000}.
\medskip

To characterize the locality of a data cloud, \cite{AR2011} recently developed a novel notion of local depth. Compared to the conventional depth notions, the most outstanding property of the local depth is its more flexibility in dealing with the applications when the underlying distributions are multimodal or have a nonconvex support. \cite{PV2013} further refined this local depth to a version that is more convenient for applications. The concept of depth-based neighborhood was also proposed, which laid the basic of many favorable inference procedures, 
such as the depth-based $k$-nearest neighbor (kNN) classifier. The depth-based kNN shares many desirable properties. For example, it is affine-equivariant and may be robust if a robust depth function is employed. The shape of the neighborhood is data-determinated. No `outside' problem exists. These consequently make the corresponding classifier very powerful in the practical data analysis \citep{PV2012}.
\medskip

All procedures here depend heavily on the concept of depth regions induced from the conventional depth notions, most of which are computationally challenging in dimensions greater than 2 nevertheless. For the case of Tukey depth, feasible algorithms have been developed by \cite{PS2012a, PS2012b} (When $p = 2$, see also \cite{RR1996}). However, these algorithms compute a Tukey depth region from the view of cutting a convex polytope with hyperplanes, and then search cone-by-cone a finite number of optimal direction vectors. To guarantee all possible cones to be taken into account, the breadth-first search algorithm is utilized in these algorithms for data of dimension $p > 2$. This practice is \emph{not so efficient}. A great proportion of computation time is spent on checking whether or not a newly obtained cone has been investigated. Furthermore, Paindaveine and \v{S}iman's approaches yield a great number of \emph{redundant} direction vectors, which result in no facet of the depth region. In practice, 
it is better to eliminate as many as possible of such direction vectors from the computation.
\medskip

In this paper, we present a new algorithm for \emph{exactly} computing a Tukey depth region for \emph{3-dimensional} data. A new tactics is utilized in order to avoid the unnecessary repeated checks as encountered when using the breadth-first search algorithm. The proposed algorithm is capable to eliminate quite a few \emph{redundant} direction vectors from considerations, and in turn save considerable computation time. The new algorithm has been efficiently implemented in \emph{Matlab}. The whole code can be obtained through emailing: \url{csuliuxh912@gmail.com} to the author; see also \emph{Appendix (A.5)}. Data examples are also provided to illustrate the performance of the proposed algorithm.
\medskip

The rest of this paper is organized as follows. Section \ref{Algorithm} provides the corresponding algorithm. 
Several data examples are given in Section \ref{Comparisons} to illustrate the performance of the proposed algorithm. Both real and simulated data are considered. Some more details are presented in the Appendix.

\vskip 0.1 in
\section{Algorithm}
\paragraph{}
\vskip 0.1 in
\label{Algorithm}

With the Tukey depth function \eqref{HDSV} at hand, a $\tau$-th Tukey depth region $\mathcal{D_{\tau}}$ is the set of all points that have at least depth $\tau$, where $0 \leq \tau \leq \tau^{*} = \sup_{x} d_{n}(x,\, F_{n})$.  That is,
\begin{eqnarray}
	\label{Dk}
	\mathcal{D}_{\tau} = \left\{x\in R^{p}: d_{n}(x,\, F_{n}) \ge \tau \right\}.
\end{eqnarray}
$\mathcal{D}_{\tau}$ is a convex polytope. The shape of $\mathcal{D}_{\tau}$ is determinated by data.
\medskip

When the observations are in general position \citep{MLB2009}, \cite{PS2011} have obtained the following lemma.
\medskip

\textbf{Lemma 1}. \textit{For $\mathcal{D}_{\tau}$ defined above, it holds that, for any $0 \leq \tau \leq \tau^{*}$, there exist a finite number $M_{1}$ of $\tau$-critical direction vectors $\mathcal{U}_{\tau} = \{u_{1},\, u_{2},\, \cdots,\, u_{M_{1}}\} \subset \mathcal{S}^{p-1}$ such that
\begin{eqnarray*}
\mathcal{D}_{\tau} = \bigcap_{u_{j} \in \mathcal{U}_{\tau}} \left\{x\in \mathcal{R}^{p} : u_{j}^{T} x \ge \tau_{u_{j}} \right\}.
\end{eqnarray*}
Here for each given $j = 1,\, 2,\, \cdots,\, M_{1}$, $u_{j} \in \mathcal{U}_{\tau}$ satisfies that: there exists at least a set of $p$ observations $\{X_{j_{1}},\, X_{j_{2}},\, \cdots,\, X_{j_{p}}\}$ such that $u_{j}$ is perpendicular to the hyperplane through these $p$ points, and  $\tau_{u_{j}} = u_{j}^{T} X_{j_{1}}$ satisfies that $\#\{i: \tau_{u_{j}} > u^{T}X_{i}\} = \lfloor n\tau \rfloor$ with $\lfloor\cdot \rfloor$ being the floor function.}
\medskip

This lemma is telling us that, to compute a $\tau$-th Tukey depth region, it is sufficient to obtain a finite number of $\tau$-critical direction vectors. Relying on this lemma, \cite{PS2012b} have developed an exact algorithm, which include the issue of computing the Tukey depth region in any dimensions as a special case. Nevertheless, this algorithm is not very computationally efficient when $p > 2$ as mentioned above, and still worthy of further improvements.
\medskip

For the special case of $p = 3$, we propose to consider the following algorithm for computing the $\tau$-critical direction vectors $\mathcal{V}_{\tau}$.
\begin{enumerate}
\item[2.1.] Set $k_{\tau} = \lfloor n \tau \rfloor + 1$, $\mathcal{A} = \text{false}(n,\, n)$, $\mathcal{T} = \text{false}(n,\, n)$\footnote{Both $\mathcal{A}$ and $\mathcal{T}$ are logic matrixes. $\mathcal{A}_{i_{0}, j_{0}} = \text{true}$ (false) means that the tuple $[i_{0},\, j_{0}]$ has (not) been considered. $\mathcal{T}_{i_{0}, j_{0}} = \text{true}$ means that the tuple $[i_{0},\, j_{0}]$ deserves further consideration because $X_{i_{0}},\, X_{j_{0}}$ have the potential to determinate a $\tau$-critical direction vector with one another observation $X_{k_{0}}$ ($k_{0} \neq i_{0},\, j_{0}$), where $\mathcal{A}_{i_{0}, j_{0}}$ and $\mathcal{T}_{i_{0}, j_{0}}$ denotes the $(i_{0},\, j_{0})$-th component of $\mathcal{A}$ and $\mathcal{T}$, respectively.}, and $\mathcal{V}_{\tau} = \emptyset$. Here $\text{false}(n,\, n)$ denotes an $n$-by-$n$ matrix of logical zeros.

\item[2.2.] Find an initial subscript tuple $[i_{0},\, j_{0}]$; \emph{see Appendix (A.1)}. Set $\mathcal{A}_{i_{0}, j_{0}} = \text{true}$ and $\mathcal{T}_{i_{0}, j_{0}} = \text{true}$. Here $i_{0}$ and $j_{0}$ should satisfy that: (a) $i_{0} > j_{0}$, (b) there exists at least one another subscript $k_{0}$ ($\neq i_{0},\, j_{0}$) such that the observations $\{X_{i_{0}},\, X_{j_{0}},\, X_{k_{0}}\}$ determinate a $\tau$-critical direction vector.

\item[2.3.] Find all the possible subscripts $k_{0} \in \mathcal{N}/\{i_{0},\, j_{0}\}$\footnote{$k_{0} \in \mathcal{N}$, but $k_{0} \notin \{i_{0},\, j_{0}\}$.} such that $\{X_{i_{0}},\, X_{j_{0}},\, X_{k_{0}}\}$ determinate a $\tau$-critical direction vector $u$; \emph{see Appendix (A.2)}. Update the set $\mathcal{V}_{\tau}$ by adding all of these $u$ into $\mathcal{V}_{\tau}$ and store the corresponding values of $u^{T} X_{i_{0}}$.

\item[2.4] For each $k_{0}$, check whether or not $\mathcal{A}_{i_{0}, k_{0}} = \text{false}$.\footnote{Here we assume $i_{0} > k_{0}$. Otherwise, replace the values of $i_{0},\, k_{0}$ with those of each other. Similarly, we assume $j_{0} > k_{0}$.} If it is, set $\mathcal{T}_{i_{0}, k_{0}} = \text{true}$. Update the value of $\mathcal{T}_{j_{0}, k_{0}}$ by using a similar procedure to $\mathcal{T}_{i_{0}, k_{0}}$. Update $\mathcal{A}$ by setting both (i) $\mathcal{A}_{i_{0}, k_{0}} = \text{true}$ and (ii) $\mathcal{A}_{j_{0}, k_{0}} = \text{true}$.

\item[2.5.] Set $\mathcal{T}_{i_{0}, j_{0}} = \text{false}$, meaning that the subscript tuple $[i_{0},\, j_{0}]$ has been investigated.

\item[2.6.] Check whether or not there is any subscript tuple $[i_{0}^{*},\, j_{0}^{*}]$ such that $\mathcal{T}_{i_{0}^{*}, j_{0}^{*}} = \text{true}$. If so, assign $[i_{0}^{*},\, j_{0}^{*}]$ to $[i_{0},\, j_{0}]$, and go back to Step 2.3. If not, eliminate the repetitions from $\mathcal{V}_{\tau}$ and terminate the algorithm successfully.
\end{enumerate}
Note that for a given 3-dimensional $\mathcal{X}^{n}$, there are ${n \choose 2}$ subscript tuples $[i_{0},\, j_{0}]$. For each $[i_{0},\, j_{0}]$, it takes $O(n\log n)$ time to compute all the possible $k_{0}$ and the critical direction vectors. Therefore, the proposed algorithm can be implemented with computational complexity at worst $O(n^{3}\log n)$ for any $\tau \in [0, \tau^{*}]$.
\medskip

This algorithm is easy to be implemented. A naive \emph{Matlab} implementation has been developed in the Appendix; \emph{see Appendix (A.5) for details}. Without loss of generality, denote $\mathcal{V}_{\tau} = \{u_{1}$, $u_{2},\, \cdots,\, u_{M_{2}}\}$ as the direction vectors computed by this algorithm, where $M_{2}$ is the number of these vectors. For $\mathcal{V}_{\tau}$, we have the following theorem; \emph{see Appendix (A.3) for its proof}.
\medskip

\textbf{Theorem 1}. \textit{Assume that the 3-dimensional observations $\mathcal{X}^{n}$ are in general position. For any $\tau \in [0,\, \tau^{*}]$, it holds that}
\begin{eqnarray*}
\mathcal{D}_{\tau} = \bigcap_{u_{j} \in \mathcal{V}_{\tau}} \left\{x\in \mathcal{R}^{3} : u_{j}^{T} x \ge \tau_{u_{j}} \right\}.
\end{eqnarray*}

Theorem 1 indicates that it is also possible to \emph{exactly} compute a $\tau$-th Tukey depth region $\mathcal{D}_{\tau}$ based on the proposed algorithm. In \emph{Matlab}, the well-developed functions such as \emph{convhulln.m} \citep{BDH1996} can be utilized to obtain all \emph{vertices or facets} of $\mathcal{D}_{\tau}$ relying on the computed $\mathcal{V}_{\tau}$ and the corresponding $\tau_{u_{j}}$'s.
\medskip

As a byproduct of Theorem 1, the following corollary may be useful in assessing the performance of the implementation of a computational algorithm; see Appendix (A.4) for its proof.
\medskip

\textbf{Corollary 1}. \textit{Assume that the 3-dimensional observations $\mathcal{X}^{n}$ are in general position. The number $M_{F}$ of the non-redundant facets of a $\tau$-th Tukey depth region $(\tau \in [0,\, \tau^{*}])$ can be upper bounded by $n (n - 1)$.}
\medskip

\noindent By the convexity property of the Tukey depth region, a critical direction vector yields at most one facet of the corresponding depth region. In this sense, Corollary 1 actually also provides an upper bound for the number of the \emph{non-redundant} $\tau$-th critical direction vectors.

\vskip 0.1 in
\section{Comparisons}
\paragraph{}
\vskip 0.1 in
\label{Comparisons}

In this section, we constructed some data examples to illustrate the performance of the proposed algorithm. All of these results are obtained on a HP Pavilion dv7 Notebook PC with Intel(R) Core(TM) i7-2670QM CPU @ 2.20GHz, RAM 6.00GB, Windows 7 Home Premium and Matlab 7.8.

\subsection{Real data}
\paragraph{}

\begin{figure}[H]
    \centering
    \captionsetup{width=0.85\textwidth}
    \includegraphics[angle=0,width=4.5in]{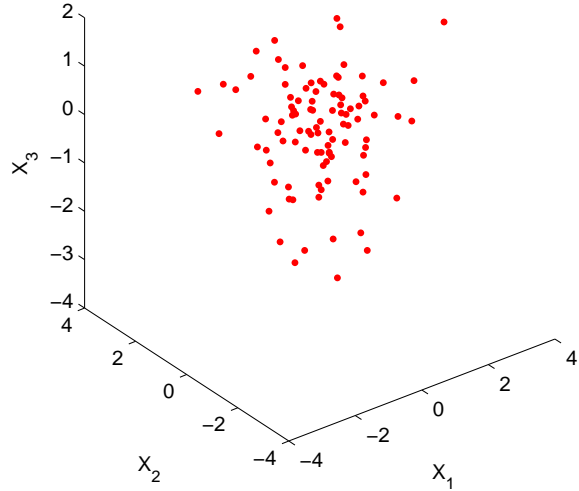}
    \caption{Shown is the scatter plot of the transformed IBM stock data from 2006 January 03 to 2006 May 25.}
    \label{fig:IBMSTA}
\end{figure}

We start with a real data set, which is a part of the the daily simple returns of IBM stock from 2006 January 03 to 2006 May 25. We use three columns under the titles of rtn, vwretd and ewretd, respectively. This data set is also used by \cite{Tsa2010}. It currently can be downloaded from his teaching page:  \url{http://faculty.chicagobooth.edu/ruey.tsay/teaching/fts3/d-ibm3dx7008.txt}. The data set consists of 100 observations. For convenience, the following transformation is performed: $Y = \widehat{\Sigma}^{-1/2} (X - \widehat{\mu})$ on the original data, where $\widehat{\mu}$, $\widehat{\Sigma}$ denote the estimated mean and covariance-matrix of the original data, respectively. The scatter plot of this transformed data set is shown in Figure \ref{fig:IBMSTA}. \emph{Remarkably}, our goal here is not to perform a thorough analysis for data, but rather to show how the algorithm works in practice.
\medskip

\begin{figure}[H]
\captionsetup{width=0.85\textwidth}
\begin{center}
    \centering
	\subfigure[$\tau = 0.01$.]{
	\includegraphics[angle=0,width=2.5in]{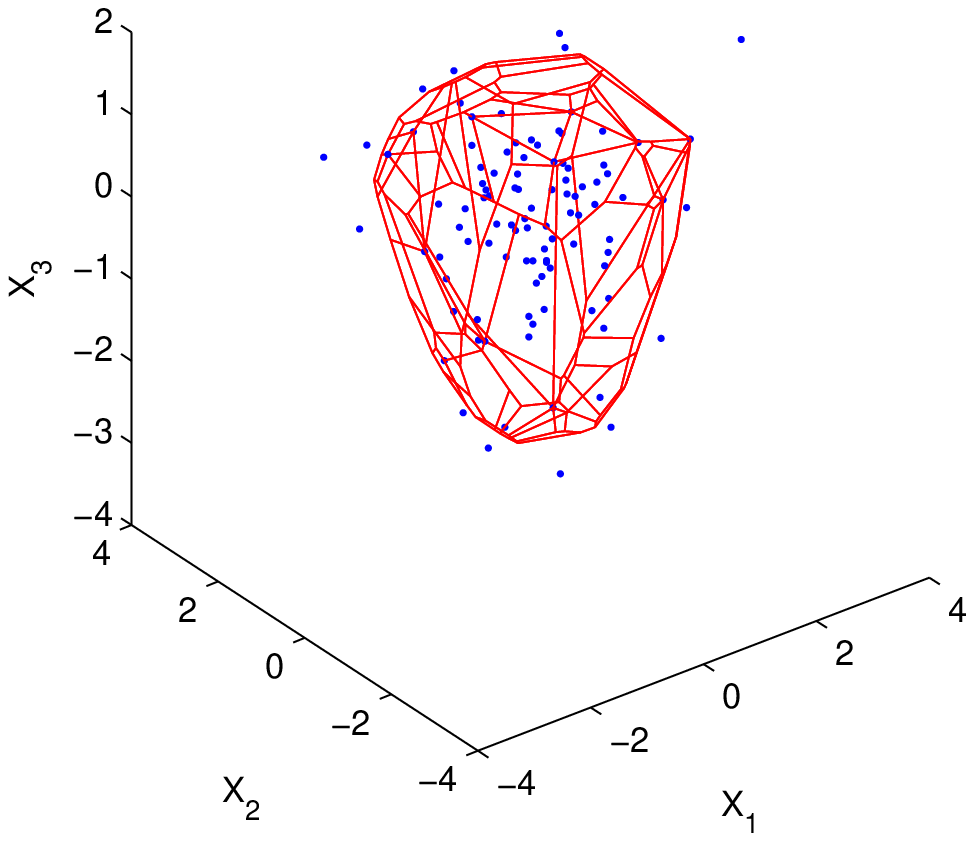}
	\label{fig:IBM001}
	}\quad
	\subfigure[$\tau = 0.05$.]{
	\includegraphics[angle=0,width=2.5in]{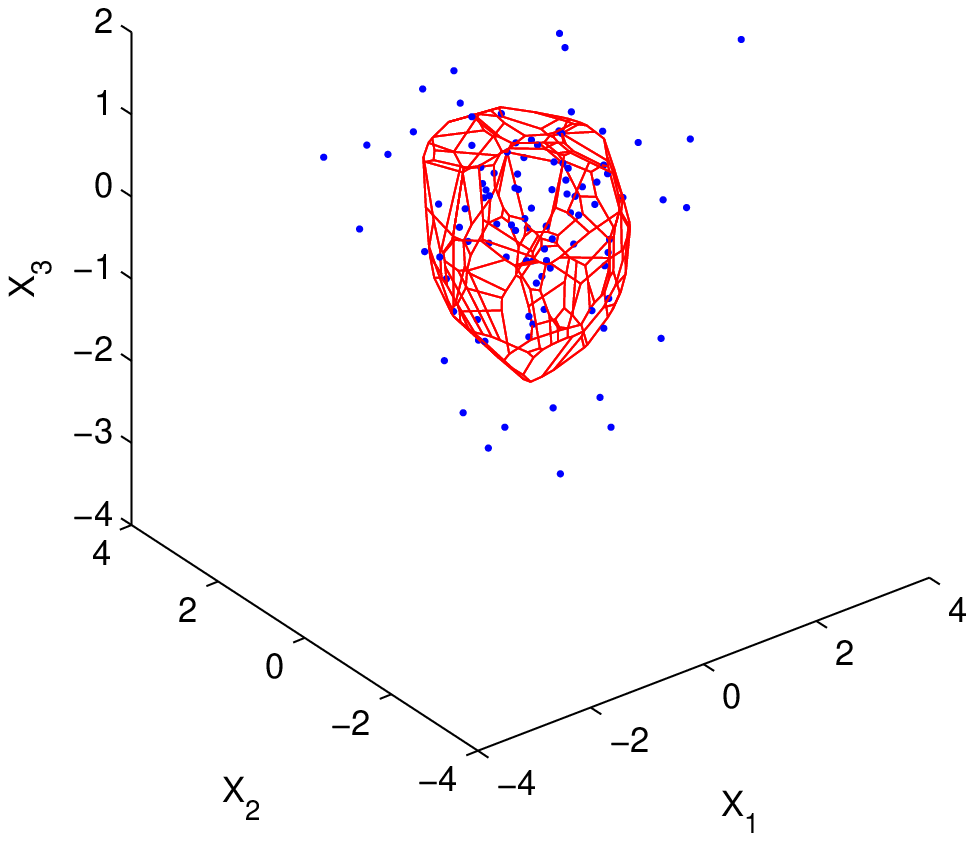}
	\label{fig:IBM005}
	}\quad
	\subfigure[$\tau = 0.10$.]{
	\includegraphics[angle=0,width=2.5in]{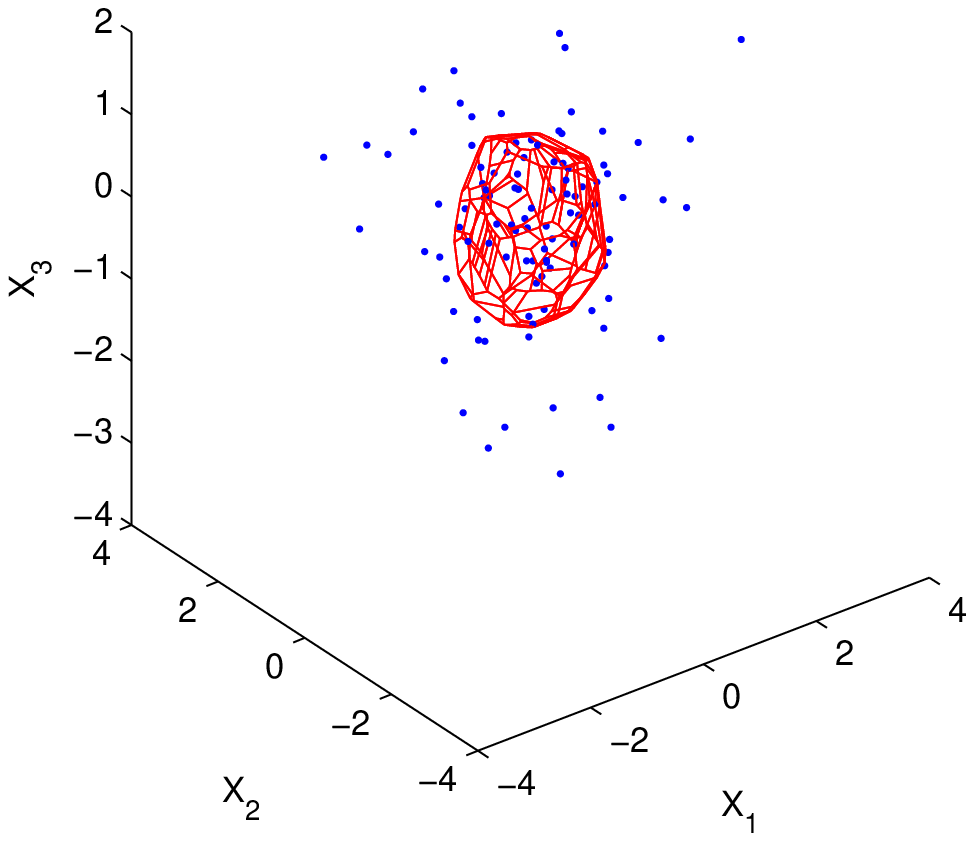}
	\label{fig:IBM010}
	}\quad
	\subfigure[$\tau = 0.20$.]{
	\includegraphics[angle=0,width=2.5in]{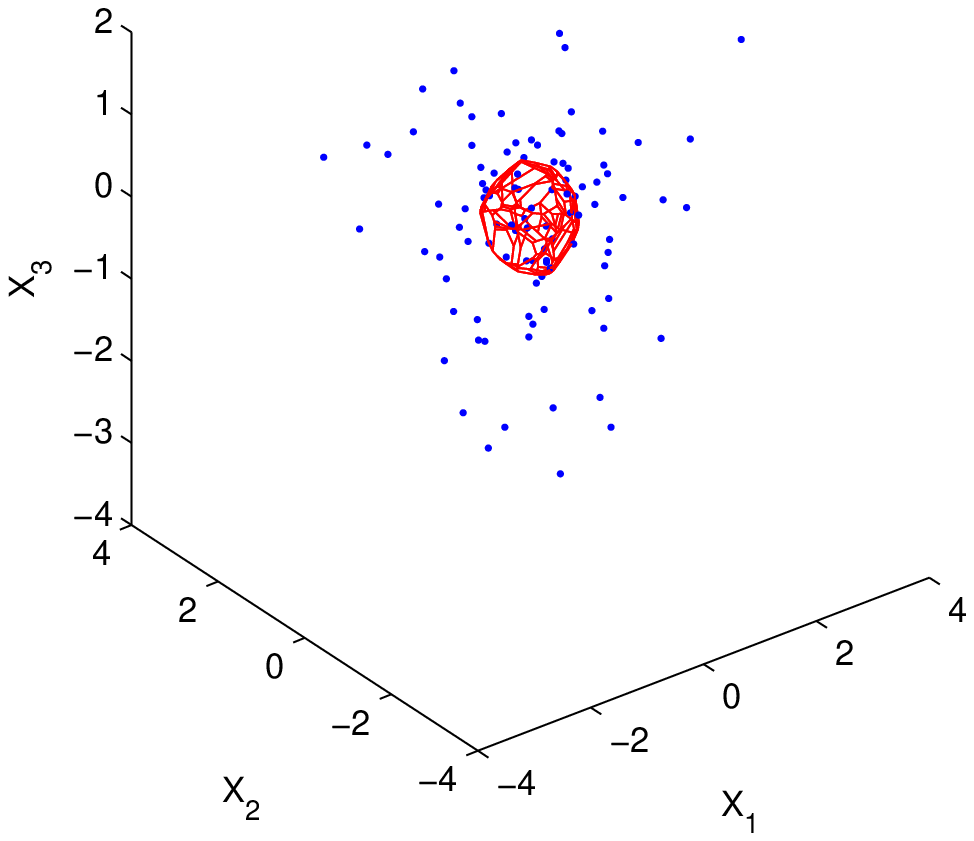}
	\label{fig:IBM020}
	}\quad
	\subfigure[$\tau = 0.30$.]{
	\includegraphics[angle=0,width=2.5in]{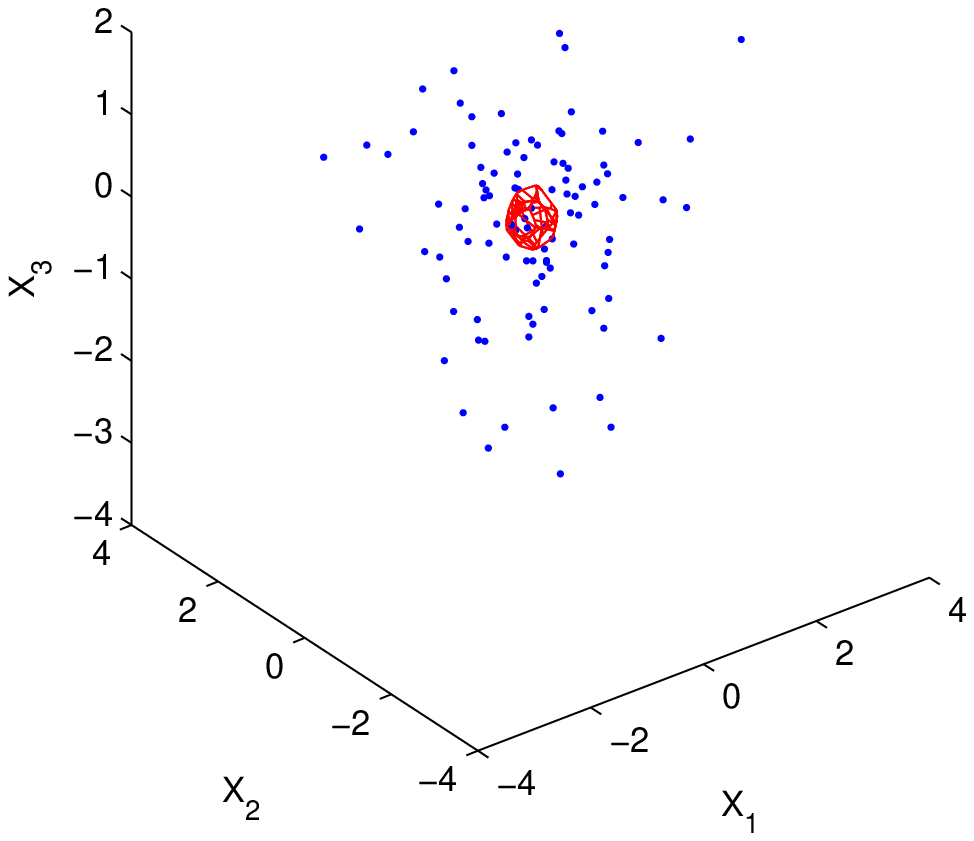}
	\label{fig:IBM030}
	}\quad
	\subfigure[$\tau = 0.35$.]{
	\includegraphics[angle=0,width=2.5in]{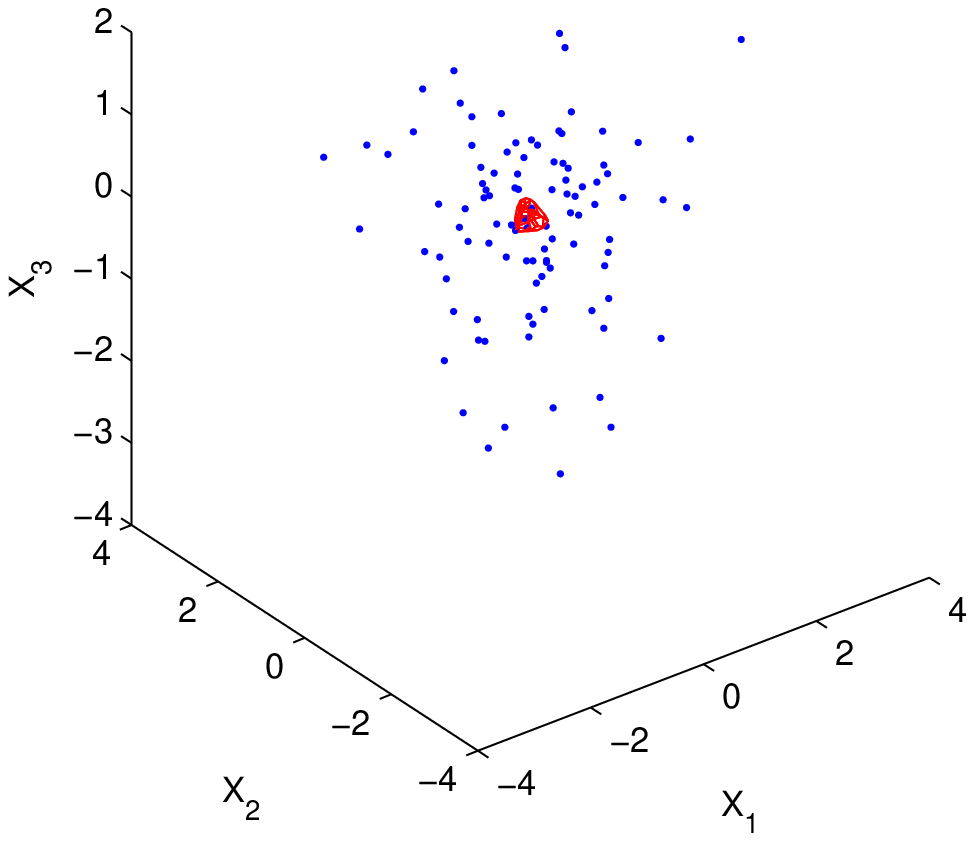}
	\label{fig:IBM035}
	}
\caption{Shown are the $0.01,\, 0.05,\, 0.10,\, 0.20,\, 0.30,\, 0.35$-th Tukey depth regions of the transformed IBM stock data from 2006 January 03 to 2006 May 25.}
\label{fig:TriHDC}
\end{center}
\end{figure}

We compute six depth regions of $\tau = 0.01$, 0.05, 0.10, 0.20, 0.30, 0.35 by using a \emph{Matlab} implementation of the proposed algorithm. It is found that the new approach yields the same results, namely, the same vertices or facets, as that (\emph{also coded in Matlab}) of \cite{PS2012b} for this data set. The results are shown in Figure \ref{fig:TriHDC}.

\begin{table}[H]
{\scriptsize
\begin{center}
    \captionsetup{width=0.85\textwidth}
    \caption{The numbers of direction vectors and the computation times (in \emph{seconds}) of the implementations of the proposed algorithm ($M_{n}$ and $T_{n}$) and that of \cite{PS2012b} ($M_{ps}$ and $T_{ps}$) for this transformed IBM stock data set.}
    \label{Tab:NumTime}
    \begin{tabular}{cp{0.5cm}p{1.0cm}p{1.0cm}p{1.0cm}rrp{1.0cm}p{1.0cm}p{1.0cm}}
    \toprule
    $\tau$\textbackslash &&\multicolumn{3}{l}{Number of direction vectors} && &\multicolumn{3}{l}{Computation times} \\
    \cmidrule(r){3-5}\cmidrule(r){8-10}
                         &&\multicolumn{1}{l}{$M_{n}$} &\multicolumn{1}{l}{$M_{ps}$} &\multicolumn{1}{l}{$M_{ps} / M_{n}$} && &\multicolumn{1}{l}{$T_{n}$} &\multicolumn{1}{l}{$T_{ps}$} &\multicolumn{1}{l}{$T_{ps} / T_{n}$} \\
    \midrule
    0.01  &&  86 &  328  &3.81 && &0.033  &0.934 &28.22\\
    0.05  && 499 & 2672  &5.35 && &0.163  &3.846 &23.56\\
    0.10  &&1195 & 6704  &5.61 && &0.410  &8.971 &21.89\\
    0.20  &&2732 &13688  &5.01 && &0.944  &17.39 &18.42\\
    0.30  &&4663 &29456  &6.32 && &1.878  &36.48 &19.43\\
    0.35  &&5106 &33768  &6.61 && &2.154  &42.34 &19.66\\
    \bottomrule
    \end{tabular}
\end{center}}
\end{table}

Furthermore, in order to gain more details about the proposed algorithm, we report the numbers of the $\tau$-th critical direction vectors obtained by the implementations of the proposed algorithm and that of \cite{PS2012b} for each depth region of this data set. It turns out that the new approach results in a much smaller number of direction vectors. For the given $\tau$, all of these numbers yielded by the proposed algorithm are smaller than the upper bound $100 \times (100 - 1)$ as suggested by Corollary 1, in contrast to many cases of the method of \cite{PS2012b}. As a result, the implementation of the proposed algorithm runs much faster than that of \cite{PS2012b}; see Table \ref{Tab:NumTime} for details. Of course, there are some limitations in the comparison. That is, we compare just the implementations, and the direction vectors computed by the method of \cite{PS2012b} may contain some repetitions. But in any case, it seems reasonable to believe that the new method outperforms that of \cite{PS2012b} for this 3-dimensional data set.

\subsection{Simulated data}
\paragraph{}

In the following, we further investigate the performance of the proposed algorithm with the simulated data, which are generated respectively from:
\begin{enumerate}
\item[] (D1). $(1 - \varepsilon) N(\textbf{0}_3, \mathbb{I}_{3}) + \varepsilon N(\textbf{0}_3, \sigma_{0}^2 \mathbb{I}_{3})$.

\item[] (D2). $(1 - \varepsilon) U([-0.5,\, 0.5]^3) + \varepsilon N(\textbf{0}_3, \sigma_{0}^2 \mathbb{I}_{3})$.

\item[] (D3). $(1 - \varepsilon) N^2(\textbf{0}_3, \Sigma_{3}) + \varepsilon N(\textbf{0}_3, \sigma_{0}^2 \mathbb{I}_{3})$.
\end{enumerate}
Here $\textbf{0}_3 = (0,\, 0,\, 0)^{T}$, $\mathbb{I}_{3}$ is the identity matrix of order 3, $\sigma_{0}^2 = 9$, $U([-0.5,\, 0.5]^3)$ denotes the 3-dimensional uniform distribution over the region $[-0.5,\, 0.5]\times [-0.5,\, 0.5]\times [-0.5,\, 0.5]$, and $N^2(\textbf{0}_3, \Sigma_{3})$ is the distribution of $X = (Z_{1}^{2},\, Z_{2}^{2},\, Z_{3}^{2})$ such that $(Z_{1},\, Z_{2},\, Z_{3})$ is subject to $N(\textbf{0}_3, \Sigma_{3})$, namely, the 3-dimensional normal distribution with mean zero and covariance-matrix
\begin{eqnarray*}
\Sigma_{3} = \left(
\begin{array}{ccccc}
1  && 0.8&& 0.8\\
0.8&& 4  && 1.6\\
0.8&& 1.6&& 4
\end{array}
\right).
\end{eqnarray*}
For any combination of $n \in \{100,\, 200,\, 300,\, 400,\, 500,\, 600\}$, $\tau \in \{$0.01, 0.05, 0.10, 0.20, 0.30$\}$ and $\varepsilon \in \{$0.00, 0.10, 0.20$\}$, we run the computation ten times for each scenario D.
\medskip

The results are listed in Table \ref{AETD1}-\ref{AETN3}. Similar to the case of real data above, the implementation of the proposed algorithm runs much faster than that of \cite{PS2012b}, and results in a much smaller number ($\leq n(n - 1)$) of direction vectors for each combination of $n$, $\tau$ and $\varepsilon$. The numbers in parentheses of these tables indicate how many times it is less than the benchmark of \cite{PS2012b}.
\medskip


\begin{table}[H]
\captionsetup{width=0.95\textwidth}
{\scriptsize
\begin{center}
    \caption{Average execution times (in seconds) of our Matlab code for Scenario (D1).}
    \label{AETD1}
    \begin{tabular}{p{0.75cm}p{0.6cm}rrrrr}
    \toprule
     \multicolumn{1}{c}{$\varepsilon$}   & \multicolumn{1}{c}{$n$}    &\multicolumn{5}{c}{$\tau$} \\
\cmidrule(r){3-7}
    &               &\multicolumn{1}{c}{0.01}  &\multicolumn{1}{c}{0.05} &\multicolumn{1}{c}{0.10}   &\multicolumn{1}{c}{0.20} &\multicolumn{1}{c}{0.30}\\
    \midrule
     0.00 &100    &0.0395 (27.23)    &0.1786 (15.56)   & 0.4329 (18.63)  &  1.1081 (17.91)  &  1.7990 (18.28)\\[0.6ex]
          &200    &0.0888 (18.99)    &0.6631 (16.47)   & 2.2206 (15.04)  &  9.0995 ( 8.94)  & 19.0790 ( 6.76)\\[0.6ex]
          &300    &0.1944 (14.50)    &2.0222 (17.45)   & 8.0355 (10.00)  & 39.0922 ( 5.21)  & 91.6904 ( 4.15)\\[0.6ex]
          &400    &0.4097 (14.47)    &4.8355 (10.26)   &21.8888 ( 6.00)  &125.9778 ( 2.74)  &296.9387 ( 1.96)\\[0.6ex]
          &500    &0.5888 (13.57)    &9.7415 ( 8.58)   &54.3214 ( 4.31)  &321.8450 ( 2.45)  &757.4869 ( 1.63)\\[0.6ex]
    0.10  &100    &0.0282 (26.49)    &0.1727 (15.69)   & 0.4107 (15.60)  &  1.0235 (18.57)  &  1.8301 (17.20)\\[0.6ex]
          &200    &0.0555 (19.57)    &0.5575 (17.76)   & 1.8831 (13.44)  &  7.8327 ( 8.74)  & 17.1017 ( 6.33)\\[0.6ex]
          &300    &0.0973 (22.51)    &1.3544 (13.12)   & 6.1476 (11.46)  & 35.5195 ( 6.10)  & 90.3325 ( 3.84)\\[0.6ex]
          &400    &0.1753 (12.02)    &3.4345 (10.29)   &17.7846 ( 6.84)  &121.7643 ( 3.54)  &328.2918 ( 2.22)\\[0.6ex]
          &500    &0.3193 ( 9.69)    &6.9518 ( 8.48)   &44.1372 ( 4.17)  &306.3422 ( 1.94)  &750.4533 ( 1.44)\\[0.6ex]
    0.20  &100    &0.0142 (52.47)    &0.1182 (17.40)   & 0.3828 (14.90)  &  0.9436 (13.96)  &  1.8105 (11.58)\\[0.6ex]
          &200    &0.0551 (32.37)    &0.4588 (15.93)   & 1.7513 (14.93)  &  8.0453 (10.35)  & 18.6599 ( 6.97)\\[0.6ex]
          &300    &0.1110 (19.18)    &1.1544 (16.82)   & 5.5883 (12.31)  & 37.7137 ( 5.75)  & 92.7185 ( 3.63)\\[0.6ex]
          &400    &0.2024 (13.14)    &2.7950 (15.16)   &15.2962 ( 8.84)  &118.5005 ( 3.37)  &305.4019 ( 2.48)\\[0.6ex]
          &500    &0.3905 (10.35)    &5.0187 ( 8.96)   &36.3917 ( 4.61)  &290.2592 ( 2.26)  &829.5661 ( 1.72)\\[0.6ex]
    \bottomrule
    \end{tabular}
\end{center}}
\end{table}

\begin{table}[H]
\captionsetup{width=0.95\textwidth}
{\scriptsize
\begin{center}
    \caption{Average numbers of the critical directions vectors obtained by our Matlab code for Scenario (D1).}
    \label{AETN1}
    \begin{tabular}{p{0.75cm}p{0.6cm}rrrrr}
    \toprule
     \multicolumn{1}{c}{$\varepsilon$}   & \multicolumn{1}{c}{$n$}    &\multicolumn{5}{c}{$\tau$} \\
\cmidrule(r){3-7}
    &               &\multicolumn{1}{c}{0.01}  &\multicolumn{1}{c}{0.05} &\multicolumn{1}{c}{0.10}   &\multicolumn{1}{c}{0.20} &\multicolumn{1}{c}{0.30}\\
    \midrule
     0.00 &100    &110 (3.27)     &  549 (3.32)     & 1333  (4.51)    &  3096  (5.03)    &  4492 (5.91)\\[0.6ex]
          &200    &229 (3.07)     & 1713 (4.86)     & 4758  (4.70)    & 11772  (5.13)    & 18454 (5.34)\\[0.6ex]
          &300    &409 (4.36)     & 3970 (6.78)     &10880  (5.69)    & 25825  (6.06)    & 40188 (6.55)\\[0.6ex]
          &400    &725 (5.73)     & 7031 (5.58)     &18559  (5.50)    & 46478  (5.43)    & 70940 (5.49)\\[0.6ex]
          &500    &841 (6.48)     &10660 (5.98)     &29139  (6.02)    & 73757  (6.79)    &111743 (6.33)\\[0.6ex]
    0.10  &100    & 86 (2.51)     &  509 (3.50)     & 1304  (3.80)    &  2968  (5.15)    &  4625 (5.55)\\[0.6ex]
          &200    &144 (2.50)     & 1536 (4.97)     & 4342  (4.69)    & 11604  (4.77)    & 17782 (4.89)\\[0.6ex]
          &300    &216 (6.26)     & 2932 (4.73)     & 9299  (6.03)    & 24891  (6.69)    & 40116 (6.44)\\[0.6ex]
          &400    &327 (3.87)     & 5573 (5.03)     &16975  (5.63)    & 46026  (6.04)    & 71964 (6.27)\\[0.6ex]
          &500    &469 (3.89)     & 8221 (5.26)     &25895  (4.95)    & 70744  (5.78)    &110547 (5.95)\\[0.6ex]
    0.20  &100    & 42 (5.14)     &  372 (3.20)     & 1209  (3.49)    &  2736  (3.74)    &  4605 (3.65)\\[0.6ex]
          &200    &145 (6.01)     & 1245 (4.23)     & 4119  (5.02)    & 11706  (5.68)    & 18558 (5.56)\\[0.6ex]
          &300    &248 (5.10)     & 2570 (5.78)     & 8632  (6.30)    & 25624  (6.48)    & 40566 (6.19)\\[0.6ex]
          &400    &368 (3.91)     & 4687 (7.06)     &15647  (6.65)    & 45278  (6.49)    & 72421 (7.00)\\[0.6ex]
          &500    &580 (4.69)     & 6615 (5.27)     &23736  (5.36)    & 69947  (5.89)    &110834 (5.72)\\[0.6ex]
    \bottomrule
    \end{tabular}
\end{center}}
\end{table}

\begin{table}[H]
\captionsetup{width=0.95\textwidth}
{\scriptsize
\begin{center}
    \caption{Average execution times (in seconds) of our Matlab code for Scenario (D2).}
    \label{AETD2}
    \begin{tabular}{p{0.75cm}p{0.6cm}rrrrr}
    \toprule
     \multicolumn{1}{c}{$\varepsilon$}   & \multicolumn{1}{c}{$n$}    &\multicolumn{5}{c}{$\tau$} \\
\cmidrule(r){3-7}
    &               &\multicolumn{1}{c}{0.01}  &\multicolumn{1}{c}{0.05} &\multicolumn{1}{c}{0.10}   &\multicolumn{1}{c}{0.20} &\multicolumn{1}{c}{0.30}\\
    \midrule
     0.00 &100    &0.6356 ( 5.09)     & 0.2476 (15.30)     & 0.5190 (14.80)    &  1.1581 (15.63)    &  1.8583 (13.08)\\[0.6ex]
          &200    &0.1370 (17.27)     & 1.0616 (12.28)     & 3.0077 (10.68)    &  9.3033 ( 9.17)    & 18.3017 ( 6.51)\\[0.6ex]
          &300    &0.3542 (12.14)     & 2.9128 (12.09)     & 9.4009 ( 8.93)    & 44.0776 ( 4.43)    & 88.5952 ( 3.30)\\[0.6ex]
          &400    &0.6518 (10.54)     & 7.7962 ( 8.69)     &30.5603 ( 4.75)    &143.1463 ( 2.48)    &313.4953 ( 1.92)\\[0.6ex]
          &500    &1.0839 ( 9.33)     &14.3471 ( 7.40)     &85.1707 ( 3.61)    &412.3910 ( 1.76)    &773.3136 ( 1.63)\\[0.6ex]
    0.10  &100    &0.0107 (43.12)     & 0.1426 (16.88)     & 0.4558 (14.35)    &  1.2381 (18.02)    &  2.0879 (21.91)\\[0.6ex]
          &200    &0.0295 (43.57)     & 0.4748 (32.00)     & 2.7920 (30.53)    &  9.9825 (21.51)    & 24.5907 (13.63)\\[0.6ex]
          &300    &0.1080 (26.98)     & 1.3357 (26.17)     & 9.3452 (19.46)    & 58.0981 ( 6.97)    &121.2798 ( 4.63)\\[0.6ex]
          &400    &0.1485 (22.31)     & 3.3645 (12.43)     &28.2102 ( 6.08)    &189.8710 ( 2.40)    &339.9063 ( 2.09)\\[0.6ex]
          &500    &0.2476 (10.17)     & 5.6117 ( 6.39)     &54.9067 ( 3.65)    &363.9800 ( 1.71)    &904.9331 ( 1.66)\\[0.6ex]
    0.20  &100    &0.0097 (49.30)     & 0.1454 (46.94)     & 0.4791 (35.52)    &  1.2875 (27.31)    &  2.2747 (20.94)\\[0.6ex]
          &200    &0.0783 (70.00)     & 0.3573 (35.06)     & 1.5052 (13.38)    & 11.8260 (12.42)    & 30.0845 ( 6.93)\\[0.6ex]
          &300    &0.1365 (21.00)     & 0.7410 (20.36)     & 7.3669 (13.85)    & 71.6315 ( 4.70)    & 99.6734 ( 4.93)\\[0.6ex]
          &400    &0.2389 ( 9.32)     & 1.5917 ( 9.59)     & 9.2006 ( 8.96)    &136.7832 ( 2.62)    &318.7450 ( 1.90)\\[0.6ex]
          &500    &0.3642 (12.77)     & 2.3915 (12.11)     &23.5790 ( 7.14)    &324.9054 ( 2.98)    &947.2455 ( 2.37)\\[0.6ex]
    \bottomrule
    \end{tabular}
\end{center}}
\end{table}

\begin{table}[H]
\captionsetup{width=0.95\textwidth}
{\scriptsize
\begin{center}
    \caption{Average numbers of the critical directions vectors obtained by our Matlab code for Scenario (D2).}
    \label{AETN2}
    \begin{tabular}{p{0.75cm}p{0.6cm}rrrrr}
    \toprule
     \multicolumn{1}{c}{$\varepsilon$}   & \multicolumn{1}{c}{$n$}    &\multicolumn{5}{c}{$\tau$} \\
\cmidrule(r){3-7}
    &               &\multicolumn{1}{c}{0.01}  &\multicolumn{1}{c}{0.05} &\multicolumn{1}{c}{0.10}   &\multicolumn{1}{c}{0.20} &\multicolumn{1}{c}{0.30}\\
    \midrule
     0.00 &100    & 158 (1.82)     &   706 (3.65)     &  1530 ( 3.87)    &   3177 ( 4.38)    &   4433 ( 4.40)\\[0.6ex]
          &200    & 348 (3.77)     &  2474 (4.04)     &  5815 ( 4.33)    &  12642 ( 5.08)    &  17635 ( 5.19)\\[0.6ex]
          &300    & 761 (4.18)     &  5311 (5.33)     & 12155 ( 5.56)    &  27603 ( 5.55)    &  39579 ( 5.45)\\[0.6ex]
          &400    &1039 (4.18)     &  9080 (5.31)     & 21581 ( 5.10)    &  48867 ( 5.12)    &  71233 ( 5.59)\\[0.6ex]
          &500    &1555 (4.75)     & 13913 (5.81)     & 34190 ( 5.84)    &  75604 ( 6.13)    & 110543 ( 6.31)\\[0.6ex]
    0.10  &100    &  28 (7.14)     &   402 (3.16)     &  1326 ( 3.23)    &   3206 ( 4.98)    &   4666 ( 6.82)\\[0.6ex]
          &200    &  72 (4.67)     &  1156 (8.62)     &  5017 (11.74)    &  12301 (11.52)    &  17999 (10.57)\\[0.6ex]
          &300    & 156 (6.87)     &  2269 (8.36)     & 10245 (10.04)    &  27082 ( 8.94)    &  41002 ( 8.43)\\[0.6ex]
          &400    & 227 (6.10)     &  4709 (5.34)     & 18677 ( 5.45)    &  48235 ( 5.44)    &  72222 ( 5.91)\\[0.6ex]
          &500    & 346 (3.82)     &  6894 (3.61)     & 28449 ( 4.84)    &  76065 ( 5.09)    & 112214 ( 5.51)\\[0.6ex]
    0.20  &100    &  24 (6.67)     &   359 (9.67)     &  1234 ( 8.08)    &   3108 ( 6.93)    &   4621 ( 5.99)\\[0.6ex]
          &200    & 134 (3.17)     &   718 (7.73)     &  2977 ( 3.69)    &  12182 ( 6.71)    &  18399 ( 6.21)\\[0.6ex]
          &300    & 198 (5.41)     &  1270 (6.03)     &  8558 ( 6.60)    &  26497 ( 7.80)    &  41445 ( 8.13)\\[0.6ex]
          &400    & 410 (2.56)     &  2802 (3.69)     & 10842 ( 5.37)    &  46060 ( 4.75)    &  74152 ( 5.38)\\[0.6ex]
          &500    & 562 (5.40)     &  3672 (6.01)     & 19369 ( 6.43)    &  74498 ( 7.74)    & 116429 ( 7.33)\\[0.6ex]
    \bottomrule
    \end{tabular}
\end{center}}
\end{table}

\begin{table}[H]
\captionsetup{width=0.95\textwidth}
{\scriptsize
\begin{center}
    \caption{Average execution times (in seconds) of our Matlab code for Scenario (D3).}
    \label{AETD3}
    \begin{tabular}{p{0.75cm}p{0.6cm}rrrrr}
    \toprule
     \multicolumn{1}{c}{$\varepsilon$}   & \multicolumn{1}{c}{$n$}    &\multicolumn{5}{c}{$\tau$} \\
\cmidrule(r){3-7}
    &               &\multicolumn{1}{c}{0.01}  &\multicolumn{1}{c}{0.05} &\multicolumn{1}{c}{0.10}   &\multicolumn{1}{c}{0.20} &\multicolumn{1}{c}{0.30}\\
    \midrule
     0.00 &100    &0.0534 (37.66)     & 0.2405 ( 5.41)     & 0.5330 (30.86)    &  1.1696 (35.98)    &  1.8622 (27.92)\\[0.6ex]
          &200    &0.1588 (27.16)     & 0.9540 (32.09)     & 3.2558 (23.56)    &  9.1112 (14.81)    & 17.2464 (10.89)\\[0.6ex]
          &300    &0.2830 (23.84)     & 2.5542 (23.08)     & 8.7608 (14.34)    & 46.8278 ( 7.25)    & 98.1411 ( 5.26)\\[0.6ex]
          &400    &0.6128 (18.36)     & 6.7065 (15.19)     &30.4956 ( 9.43)    &136.3922 ( 4.55)    &275.2976 ( 3.61)\\[0.6ex]
          &500    &1.1261 (15.00)     &13.9530 (11.39)     &72.2767 ( 6.24)    &361.9681 ( 3.63)    &721.6153 ( 2.97)\\[0.6ex]
    0.10  &100    &0.0210 (48.32)     & 0.1645 (37.49)     & 0.4441 (36.57)    &  1.3561 (31.72)    &  2.1059 (24.11)\\[0.6ex]
          &200    &0.0593 (33.34)     & 0.7486 (31.92)     & 2.8938 (29.51)    &  9.7552 (17.97)    & 18.2607 (12.44)\\[0.6ex]
          &300    &0.1063 (19.33)     & 1.4610 (18.87)     & 7.8851 (16.40)    & 42.3759 ( 7.43)    & 91.0750 ( 5.02)\\[0.6ex]
          &400    &0.2257 (16.76)     & 2.5603 (14.12)     &21.8409 (15.81)    &119.8576 ( 6.80)    &304.2660 ( 4.12)\\[0.6ex]
          &500    &0.3739 (13.30)     & 4.8199 (12.03)     &43.9444 (10.74)    &312.6973 ( 5.15)    &777.7728 ( 3.45)\\[0.6ex]
    0.20  &100    &0.0324 (49.54)     & 0.1342 (33.42)     & 0.3001 (33.00)    &  1.0985 (43.20)    &  1.9382 (33.24)\\[0.6ex]
          &200    &0.0787 (27.05)     & 0.5317 (22.81)     & 1.9892 (32.72)    &  9.0659 (18.86)    & 18.6073 (13.36)\\[0.6ex]
          &300    &0.1234 (23.46)     & 1.1977 (19.86)     & 5.3343 (15.52)    & 39.0659 ( 9.93)    & 95.4297 ( 6.21)\\[0.6ex]
          &400    &0.2734 (24.97)     & 2.4064 (15.15)     &10.4868 ( 7.88)    &115.3662 ( 5.63)    &380.4795 ( 3.54)\\[0.6ex]
          &500    &0.5063 (12.92)     & 5.9210 (10.24)     &31.7895 ( 8.85)    &335.6343 ( 4.56)    &790.5407 ( 3.97)\\[0.6ex]
    \bottomrule
    \end{tabular}
\end{center}}
\end{table}

\begin{table}[H]
\captionsetup{width=0.95\textwidth}
{\scriptsize
\begin{center}
    \caption{Average numbers of the critical directions vectors obtained by our Matlab code for Scenario (D3).}
    \label{AETN3}
    \begin{tabular}{p{0.75cm}p{0.6cm}rrrrr}
    \toprule
     \multicolumn{1}{c}{$\varepsilon$}   & \multicolumn{1}{c}{$n$}    &\multicolumn{5}{c}{$\tau$} \\
\cmidrule(r){3-7}
    &               &\multicolumn{1}{c}{0.01}  &\multicolumn{1}{c}{0.05} &\multicolumn{1}{c}{0.10}   &\multicolumn{1}{c}{0.20} &\multicolumn{1}{c}{0.30}\\
    \midrule
     0.00 &100    & 213 ( 4.06)     &  785 (5.23)     & 1666 ( 7.34)    & 3445 ( 8.26)    &  4894 ( 7.77)\\[0.6ex]
          &200    & 603 ( 3.79)     & 3222 (6.57)     & 7012 ( 7.55)    &13625 ( 6.99)    & 18968 ( 6.85)\\[0.6ex]
          &300    & 929 ( 4.63)     & 5703 (7.00)     &12974 ( 7.06)    &29113 ( 7.60)    & 42981 ( 7.68)\\[0.6ex]
          &400    &1649 ( 4.55)     &11528 (6.40)     &26891 ( 7.25)    &53417 ( 7.28)    & 73666 ( 7.13)\\[0.6ex]
          &500    &2286 ( 4.89)     &16515 (6.43)     &38561 ( 7.20)    &81172 ( 7.92)    &114403 ( 7.85)\\[0.6ex]
    0.10  &100    &  59 ( 4.61)     &  547 (6.93)     & 1406 ( 8.03)    & 3401 ( 8.32)    &  4793 ( 6.98)\\[0.6ex]
          &200    & 132 ( 4.67)     & 2030 (7.36)     & 5995 ( 9.26)    &13194 ( 9.13)    & 19061 ( 7.98)\\[0.6ex]
          &300    & 225 ( 4.98)     & 3340 (5.83)     &12077 ( 7.71)    &29041 ( 7.43)    & 41614 ( 7.18)\\[0.6ex]
          &400    & 393 ( 5.72)     & 4145 (5.93)     &19838 (11.40)    &48840 ( 9.62)    & 75828 ( 8.34)\\[0.6ex]
          &500    & 549 ( 5.79)     & 6351 (6.68)     &29920 (10.23)    &79100 ( 9.51)    &119544 ( 8.70)\\[0.6ex]
    0.20  &100    &  90 ( 8.62)     &  407 (7.25)     &  890 ( 8.09)    & 3208 (11.07)    &  4848 (10.16)\\[0.6ex]
          &200    & 191 ( 6.03)     & 1361 (6.46)     & 4915 ( 9.79)    &13290 ( 9.31)    & 19472 ( 9.03)\\[0.6ex]
          &300    & 259 ( 7.01)     & 2487 (7.23)     & 8245 ( 7.46)    &27363 ( 9.63)    & 43011 ( 8.69)\\[0.6ex]
          &400    & 459 (10.16)     & 3970 (6.91)     &12053 ( 5.07)    &46801 ( 8.39)    & 75364 ( 8.60)\\[0.6ex]
          &500    & 686 ( 6.02)     & 6970 (6.24)     &22555 ( 8.22)    &75553 ( 9.21)    &117714 ( 8.17)\\[0.6ex]
    \bottomrule
    \end{tabular}
\end{center}}
\end{table}

\vskip 0.1 in
\section{Concluding remarks}
\paragraph{}
\vskip 0.1 in


In this paper, we have constructed a fast algorithm for computing a 3-dimensional $\tau$-Tukey depth region. Rather than searching the critical direction vectors cone-by-cone, the proposed algorithm finds all possible direction vectors subscript-tuple-by-subscript-tuple. 
Consequently, checking directly the values of $\mathcal{A}_{i_{0}, j_{0}}$ and $\mathcal{T}_{i_{0}, j_{0}}$ is sufficient to determine if a newly obtained subscript tuple $[i_{0},\, j_{0}]$ has been investigated. This new searching tactics helps to avoid some unnecessary repeated checks and in turn save considerable computational times. The data examples indicate that our results provide a significant speed-up over existing algorithms. 
\medskip

In the literature, there are many other depth notions, such as projection depth and zonoid depth, closely related to the methodology of projection pursuit. It turns out that most of them can be exactly computed from the view of cutting a convex polytope with hyperplanes; see \cite{MLB2009} and \cite{LZ2014} respectively for details. Then a natural question concerns faster algorithms for these depth notions. This may be of great practice interest, because some of these depth notions could not be computed efficiently in dimensions of $p\ge 3$. Work is underway.






\vskip 0.1 in
\section*{Acknowledgments}
\paragraph{}
\vskip 0.1 in


This research is partly supported by the National Natural Science Foundation of China (Grant No.11361026, No.11161022, No.61263014), and the Natural Science Foundation of Jiangxi Province (Grant No.20122BAB201023, No.20132BAB201011).

\vskip 0.1 in
\section*{{\sc Appendix}}
\vskip 0.1 in

\noindent \textit{$\square$ (A.1) Find an initial subscript tuple $[i_{0},\, j_{0}]$}. In Step 2.2, we compute $[i_{0},\, j_{0}]$ by using the following procedure.
\begin{enumerate}
\item[] 2.2.1. Generate a random unit vector $u_{0}$, and store the permutation $(i_{1},\, i_{2},\, \cdots,\, i_{n})$ such that $u_{0}^{T}X_{i_{1}} < u_{0}^{T}X_{i_{2}} < \cdots < u_{0}^{T}X_{i_{n}}$.

\item[] 2.2.2. Compute the distances $\theta_{i}$'s between the point $u_{0}$ and $n - 1$ hyperplanes $\mathcal{P}_{i} = \{s \in \mathcal{R}^{p}: (X_{i} - X_{i_{k_{\tau}}})^T s = 0\}$, where $i \in \mathcal{N}$ and $i \neq i_{k_{\tau}}$.

\item[] 2.2.3. Find the minimum among $\theta_{i}$'s and obtain the corresponding subscript tuple $[i^{*},\, i_{k_{\tau}}]$. Assign the maximum of $\{i^{*},\, i_{k_{\tau}}\}$ to $i_{0}$ and the other one to $j_{0}$, respectively.
\end{enumerate}

This procedure corresponds to the code snippets between lines 48-58 of \emph{FHC3D.m}; see \emph{Appendix (A.5)}. The rational behind is as follows. By $u_{0}^{T}X_{i_{1}} < u_{0}^{T}X_{i_{2}} < \cdots < u_{0}^{T}X_{i_{n}}$, it is easy to show that $u_{0} \in \mathcal{C} = \{t: \mathbb{A}_{0}^{T} t \leq 0\}$, where
\begin{eqnarray*}
	\mathbb{A}_{0} = (X_{i_{1}} - X_{i_{k_{\tau}}},\, X_{i_{2}} - X_{i_{k_{\tau}}},\,\cdots,\, X_{i_{k_{\tau} - 1}} - X_{i_{k_{\tau}}},\,  X_{i_{k_{\tau}}} - X_{i_{k_{\tau} + 1}},\, \cdots,\, X_{i_{k_{\tau}}} - X_{i_{n}}).
\end{eqnarray*}
Clearly, $\mathcal{C}$ forms a polytope, on each vertex of which must lie an $\tau$-critical direction vectors. The closest hyperplane to $u_{0}$ must pass through a \emph{non-redundant} facet of $\mathcal{C}$, and hence its corresponding subscript tuple $[i^{*},\, i_{k_{\tau}}]$ is what we want.
\endp
\bigskip

\noindent\textit{$\square$ (A.2) Find all the possible subscripts $k_{0}$}. In Step 2.3, we utilize the following procedure to to find all the possible subscripts $k_{0}$.
\begin{enumerate}
\item[] 2.3.k1. Project the data points $X_{k}$ $(k \in \mathcal{N}/\{i_{0},\, j_{0}\})$ onto the plane $\mathcal{P}_{0}$, which is perpendicular to $\alpha_{0} = \frac{X_{i_{0}} - X_{j_{0}} }{\|X_{i_{0}} - X_{j_{0}}\|}$ and pass through $X_{j_{0}}$. Without loss of generality, denote the projection of $X_{k}$ as $X_{k}^{*}$.

\item[] 2.3.k2. Compute the polar coordinate angles $\theta_{k}$ $(\theta_{k} \in [-\pi, \, \pi))$ of $(\beta_{k1},\, \beta_{k2})$ if $\alpha_{03} \neq 0$ (otherwise, use $(\beta_{k2},\, \beta_{k3})$ instead of $(\beta_{k1},\, \beta_{k2})$), where $\beta_{k} := (\beta_{k1},\, \beta_{k2},\, \beta_{k3})^{T} = \frac{X_{k}^{*} - X_{j_{0}}}{\|X_{k}^{*} - X_{j_{0}}\|}$ $(k \in \mathcal{N}/\{i_{0},\, j_{0}\})$ and $\alpha_{03}$ denotes the third component of $\alpha_{0}$.

\item[] 2.3.k3. For each $k \in \mathcal{N}/\{i_{0},\, j_{0}\}$, count the number $N_{1}$ of these polar coordinate angles that lie in $(\theta_{k},\, \theta_{k} + \pi)$ and the number $N_{2}$ of those that lie in $(-\pi,\, \theta_{k})\bigcup (\theta_{k} + \pi,\, \pi)$. If either $N_{1}$ or $N_{2}$ is equal to $\lfloor n\tau \rfloor$, then $k$ is a satisfactory subscript.

\end{enumerate}

\begin{figure}[H]
\captionsetup{width=0.85\textwidth}
\centering
\includegraphics[angle=0,width=6.0in]{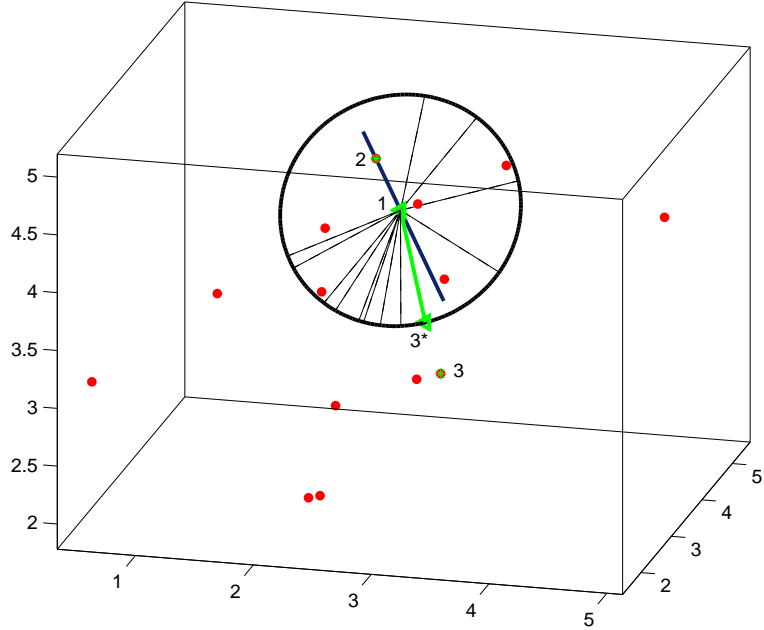}
\caption{Shown is an illustration of how to find the possible subscripts $k_{0}$. The points denote the observations. Every point (not in the line passing through points 1 and 2) corresponds to a unit vector stemming from point 1 in $\mathcal{P}_{0}$.}
\label{fig:A2}
\end{figure}

This part can be easily implemented by using the Gram-Schmidt orthonormalization. The corresponding \emph{Matlab} code lie between lines 67-87 of \emph{FHC3D.m}; see \emph{Appendix (A.5)}. As illustration of this procedure is provided in Figure \ref{fig:A2}. In this figure, points 1 and 2 serve as the points $X_{j_{0}}$ and $X_{i_{0}}$, respectively. Then every point $X_{k}$ $(k \in \mathcal{N}/\{i_{0},\, j_{0}\})$ corresponds to the polar coordinate angle of one unit vector in $\mathcal{P}_{0}$ passing through point 1. For example, $\alpha_{0}$ lies in the line pass through points 1 and 2, and point 3 corresponds to the angle of the vector connecting points 1 and $3^{*}$. It is easy to see that the plane passing through points 1, 2 and 3 divides the whole space $\mathcal{R}^{3}$ into two halfspace spaces with 4 data points on one side, and 12 data points on the other side. A similar procedure of such kind is the planar algorithm developed by \cite{RS1998} (pp. 201-202). \endp
\bigskip

\noindent\textit{$\square$ (A.3) Proof of Theorem 1}. For a given $\tau$, let $k_{\tau} = \lfloor n \tau \rfloor + 1$. Note that for any $u_{0} \in \mathcal{S}^{p-1}$, there must exist a permutation $(j_{1},\, j_{2},\,\cdots,\, j_{n})$ of $(1,\, 2,\, \cdots,\, n)$ such that $u^{T}X_{j_{1}} < u^{T}X_{j_{2}} < \cdots < u^{T}X_{j_{n}}$. Using this, one can, similar to \cite{LZW2013}, obtain that
\begin{eqnarray*}
    \mathcal{S}^{p-1} = \bigcup_{l=1}^{M_{s}} \mathcal{S}_{l}, \quad \text{with} \quad \mathcal{S}_{l}=\{u\in \mathcal{S}^{p-1}: A_{l}^{T}u \leq 0\},
\end{eqnarray*}
where $M_{s}$ denotes the number of $\mathcal{S}_{l}$ and
\begin{eqnarray*}
	\mathbb{A}_{l} = (X_{j_{l,1}} - X_{j_{l,k_{\tau}}},\, X_{j_{l,2}} - X_{j_{l,k_{\tau}}},\,\cdots,\, X_{j_{l,k_{\tau} - 1}} - X_{j_{l,k_{\tau}}},\,  X_{j_{l,k_{\tau}}} - X_{j_{l,k_{\tau} + 1}},\, \cdots,\, X_{j_{l,k_{\tau}}} - X_{j_{l,n}}).
\end{eqnarray*}

Denote $\mathcal{C}_{l} = \{t\in \mathcal{R}^{p}: A_{l}^{T}t \leq 0\}$ ($1\leq l \leq M_{s}$). Clearly, $\mathcal{S}_{l} \subset \mathcal{C}_{l}$ and $\mathcal{C}_{l}$'s are convex cones. Without loss of generality, assume $\mathcal{C}_{l}$ has $m_{l}$ vertices, and let $\widetilde{u}_{l,1},\, \widetilde{u}_{l,2},\, \cdots,\, \widetilde{u}_{l,m_{l}}\ (\in \mathcal{S}_{l} \cap \mathcal{C}_{l})$ to be the unit direction vectors corresponding to these vertices. By the convexity of $\mathcal{C}_{l}$ and the fact that $\widetilde{u}_{l,1}^{T}x \ge \widetilde{u}_{l,1}^{T}X_{j_{l,k_{\tau}}}$, $\cdots$, $\widetilde{u}_{l, m_{l}}^{T}x \ge \widetilde{u}_{l, m_{l}}^{T}X_{j_{l,k_{\tau}}}$ together lead to $(\sum_{i=1}^{m_{l}}\lambda_{i} \widetilde{u}_{l,i})^{T}x \ge (\sum_{i=1}^{m_{l}}\lambda_{i} \widetilde{u}_{l,i})^{T}X_{j_{l,k_{\tau}}}$, it is easy to show that
\begin{eqnarray*}
    \bigcap_{u\in \mathcal{C}_{l}} \left\{x\in \mathcal{R}^{p}: u^{T} x \ge u^{T} X_{j_{l,k_{\tau}}}\right\} = \bigcap_{u\in \mathcal{S}_{l}} \left\{x\in \mathcal{R}^{p}: u^{T} x \ge u^{T} X_{j_{l,k_{\tau}}}\right\} = \bigcap_{i = 1}^{m_{l}} \left\{x\in \mathcal{R}^{p}: \widetilde{u}_{l,i}^{T} x \ge \widetilde{u}_{l,i}^{T} X_{j_{l,k_{\tau}}}\right\},
\end{eqnarray*}
where $\lambda_{i} \ge 0$, $i=1,\, 2,\, \cdots,\, m_{l}$. 
This implies that the exact computation of $\mathcal{D}_{\tau}$ depends only on a finite number of unit direction vectors corresponding to the vertices of $\mathcal{C}_{l}$, $l = 1,\,\cdots,\, M_{s}$.

\begin{figure}[H]
\captionsetup{width=0.85\textwidth}
\centering
\includegraphics[angle=0,width=5.0in]{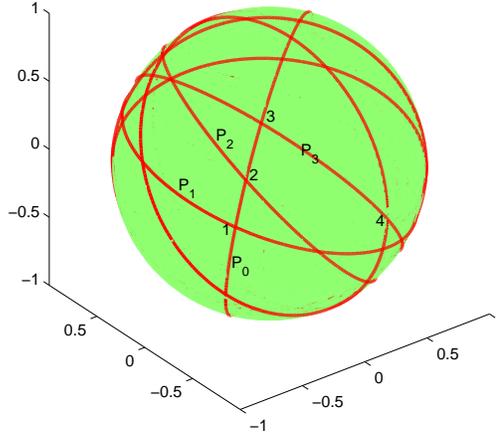}
\caption{Shown is an illustration of the linked points on the unit sphere $\mathcal{S}^{2}$. Here points 2 and 4 are linked through the arcs between points 2, 3 and 4.}
\label{fig:Th1}
\end{figure}

When $p = 3$, a vertex of $\mathcal{C}_{l}$ is determined by two non-redundant facets, which are determined by three observations. Every two points, corresponding to two critical direction vectors, on the sphere $\mathcal{S}^{2}$ are linked with each other through some arcs if the observations are in general position; see points 1 and 4 in Figure \ref{fig:Th1} for an illustration. A subscript tuple, corresponding to two observations, determines a non-redundant facet, which may contain several critical direction vectors. Enumerating all such subscript tuples, namely, iterating Steps 2.3-2.6, can find the critical direction vectors, by using which it is sufficient to obtain an exact Tukey depth region. \endp

\begin{figure}[H]
\begin{center}
    \centering
	\subfigure[]{
	\includegraphics[angle=0,width=5in]{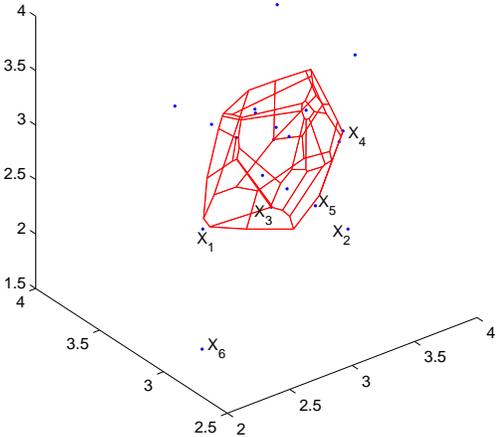}
	\label{fig:HC1}
	}\quad
	\subfigure[]{
	\includegraphics[angle=0,width=5in]{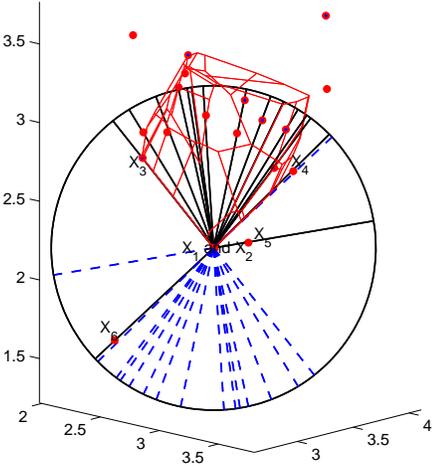}
	\label{fig:HC2}
	}
\caption{\ref{fig:HC1} is the $0.01$-th Tukey depth region of a data set with sample size $n=20$. For this case, $k_{\tau} = 1$, and there are four $\tau$-critical hyperplanes passing through $X_{1}$, $X_{2}$ (and $X_{3}$, $X_{4}$, $X_{5}$, $X_{6}$, respectively). Among them, only two hyperplanes, through $\{X_{1}$, $X_{2}$, $X_{3}\}$ and $\{X_{1}$, $X_{2}$, $X_{4}\}$ respectively, yield two $non$-redundant facets by the convexity of $\mathcal{D}_{\tau}$; see \ref{fig:HC2}.}
\label{fig:HC12}
\end{center}
\end{figure}

\noindent\textit{$\square$ (A.4) Proof of Corollary 1}. Without loss of generality, we call a hyperplane $\tau$-critical hyperplane if it passes though three observations and divides the whole space into two parts with $\lfloor n \tau \rfloor$ on one side and the rest on the other side. By the convexity of the Tukey depth region, a $\tau$-critical hyperplane yields at most one facet of $\mathcal{D}_{\tau}$. When $p = 3$ and the observations are in general position, although every two observations $X_{i_{0}}$ and $X_{j_{0}}$, corresponding to $[i_{0},\, j_{0}]$, may be contained in more than two $\tau$-critical hyperplanes, at most two of these $\tau$-critical hyperplanes are possible to yield $non$-redundant facets of $\mathcal{D}_{\tau}$; see Figure \ref{fig:HC12} for an illustration. The proves that the number of $non$-redundant facets of $\mathcal{D}_{\tau}$ is at most $2 \times {n \choose 2} = n (n - 1)$. \endp
\bigskip

\noindent\textit{$\square$ (A.5) Code snippet}. The main function \emph{FHC3D.m} corresponding to the proposed algorithm. It is construed mainly for computing the $\tau$-critical direction vectors of a given Tukey depth region.
\begin{lstlisting}
% FHC3D.m
%
%   Description:
%
%      Fast computing the critical direction vectors of the tau-th
%      Tukey depth region for a 3-dimensional X.
%
% Input arguments:
%               X            - data matrix, n-by-3 (matrix), n denotes the
%                              sample size
%            tau0            - depth value of the computed contour
%                              (0 <= tau0 < tau0Star (the maximum depth
%                              value))
%
% Output arguments:
%            vecu            - struct('u', [], 'QuanV', [], 'NumU', [])


function vecu = FHC3D(X, tau0)

    % Check the input arguments
    [n, p] = size(X);
    if p ~= 3, error('X must be an n-by-3 matrix!'); end

    % Initialize vecu
    vecu.u = [];    vecu.QuanV = [];    vecu.NumU = 0;

    % Initialize the archives NewIndx and OldIndx.
    % 'NewIndx(i, j) = true' means that Xi - Xj needs to be considered;
    % 'OldIndx(i, j) = true' means that Xi - Xj has been considered.
    NewIndx = false(n, n);    OldIndx = false(n, n);

    % The sub-index tauk corresponding to tau0
    taukSUB1 = floor(n * tau0);    tauk = taukSUB1 + 1;

    % Initialize some intermediate variables
    nSUB1     = n - 1;                 nSUB2     = n - 2;
    nDIV4     = floor(n / 4);
    ONESn1X1  = ones(nSUB1, 1);        ONESn2X1  = ones(nSUB2, 1);
    ONESpX1   = ones(p, 1);            piMULT2   = 2 * pi;
    nSubtauk1 = nSUB2 - taukSUB1;      nSubtauk2 = nSubtauk1 - 1;
    VecN1     = 1:nSUB2;
    LowIndx1  = VecN1 + taukSUB1;       UpIndx1  = VecN1 + tauk;
    LowIndx2  = VecN1 + nSubtauk2;      UpIndx2  = VecN1 + nSubtauk1;
    LowIndx3  = LowIndx2 - nDIV4;       UpIndx3  = UpIndx1 + nDIV4;

    % Obtain an intial index-couple [rowi, colj]
    IndxSet = [1:(tauk - 1), (tauk + 1):n];
    IPVec   = ONESpX1 / norm(ONESpX1);
    [sortXu, perm0]      = sort(X * IPVec);
    QuanX                = X(perm0, :);
    % Obtain the normal vectors of {u: u^T * (X(i, :) - X(j, :))}
    NVec                 = (QuanX(IndxSet, :) - ONESn1X1 * QuanX(tauk, :));
    NVec(tauk:nSUB1, :)  = -NVec(tauk:nSUB1, :);
    NVec                 = NVec ./ (sqrt(sum(NVec.^2, 2)) * ONESpX1');
    [tmpv, tmpk]         = min(abs(NVec * IPVec));
    rowi = max(perm0(IndxSet(tmpk(1))), perm0(tauk));
    colj = min(perm0(IndxSet(tmpk(1))), perm0(tauk));

    % Update NewIndx and OldIndx
    NewIndx(rowi, colj) = true;
    OldIndx(rowi, colj) = true;

    % Compute all the optimal direction vectors
    while any(any(NewIndx)) % If NewIndx(i, j) = true
        [rowi, colj] = find(NewIndx, 1);
        IndxSet = [1:(colj - 1), (colj + 1):(rowi - 1), (rowi + 1):n];
        % The vector Xi - Xj
        alpha0 = X(rowi, :) - X(colj, :);
        alpha0 = alpha0 / norm(alpha0);
        % The vectors Xk - Xj (k in {1, 2, ..., n} - {i, j})
        beta0  = X(IndxSet, :) - ONESn2X1 * X(colj, :);
        gamma0 = beta0 - (beta0 * alpha0') * alpha0;
        tmpvec = gamma0(:, 1);
        if abs(alpha0(3)) < 1e-12, tmpvec = gamma0(:, 3); end
        isvec0 = (tmpvec < 0);
        theta0 = atan(gamma0(:, 2) ./ tmpvec) + ...
            (isvec0 & (gamma0(:, 2) > 0)) * pi - ...
            (isvec0 & (gamma0(:, 2) < 0)) * pi;
        % Sort theta0
        [theta0, perm0] = sort(theta0);
        theta1 = [theta0; theta0(1:nSubtauk1) + piMULT2];
        perm1  = [perm0; perm0(1:nSubtauk1)];

        UpBnd = theta0 + pi;
        isvec1 = (theta1(LowIndx1) < UpBnd) & (UpBnd < theta1(UpIndx1));
        isvec2 = (theta1(LowIndx2) < UpBnd) & (UpBnd < theta1(UpIndx2));

        % Update vecu
        newl1 = perm1(UpIndx3(isvec1));
        newl2 = perm1(VecN1(isvec1));
        for ll = 1:length(newl1)
            tmpu0 = gamma0(newl2(ll), :) / norm(gamma0(newl2(ll), :));
            tmpu1  = gamma0(newl1(ll), :) - (gamma0(newl1(ll), :) * tmpu0') * tmpu0;
            tmpu1 = tmpu1' / norm(tmpu1);
            tmpv = X(rowi, :) * tmpu1;
            vecu.u     = [vecu.u, tmpu1];
            vecu.QuanV = [vecu.QuanV; tmpv];
            vecu.NumU  = vecu.NumU + 1;
        end
        newl1 = perm1(LowIndx3(isvec2));
        newl2 = perm1(VecN1(isvec2));
        for ll = 1:length(newl1)
            tmpu0 = gamma0(newl2(ll), :) / norm(gamma0(newl2(ll), :));
            tmpu1  = gamma0(newl1(ll), :) - (gamma0(newl1(ll), :) * tmpu0') * tmpu0;
            tmpu1 = tmpu1' / norm(tmpu1);
            tmpv = X(rowi, :) * tmpu1;
            vecu.u     = [vecu.u, tmpu1];
            vecu.QuanV = [vecu.QuanV; tmpv];
            vecu.NumU  = vecu.NumU + 1;
        end

        % Update NewIndx and OldIndx
        newl2 = perm0(isvec1 | isvec2);
        for ll = 1:length(newl2)
            tmpv1 = sort([rowi, IndxSet(newl2(ll))], 'descend');
            tmpv2 = sort([colj, IndxSet(newl2(ll))], 'descend');
            if ~OldIndx(tmpv1(1), tmpv1(2)), NewIndx(tmpv1(1), tmpv1(2)) = true; end
            if ~OldIndx(tmpv2(1), tmpv2(2)), NewIndx(tmpv2(1), tmpv2(2)) = true; end
            OldIndx(tmpv1(1), tmpv1(2)) = true;
            OldIndx(tmpv2(1), tmpv2(2)) = true;
        end

        % Eliminate NewIndx(rowi, colj) from the next consideration
        NewIndx(rowi, colj) = false;
    end
    % Eliminate the repetitions from vecu
    [tmpv, indx]   = unique(num2str(vecu.u', 12), 'rows');
    vecu.u         = vecu.u(:, indx);
    vecu.QuanV     = vecu.QuanV(indx);
    vecu.NumU      = length(indx);

% End of program
\end{lstlisting}
\endp

\end{document}